\documentclass[12pt]{iopart}
\usepackage{iopams}  
\expandafter\let\csname equation*\endcsname\relax

\expandafter\let\csname endequation*\endcsname\relax

\usepackage{color}
\usepackage{graphicx}
\usepackage{amsmath}

\begin{document}

\title[Supersymmetric Quantum Electronic States in Graphene Under Uniaxial Strain]{Supersymmetric Quantum Electronic States in Graphene Under Uniaxial Strain}

\author{Y.~Concha$^1$, A.~Huet$^2$, A. Raya$^3$ and D.~Valenzuela$^4$}

\address{
$^1$ Facultad de Ingenier\'{\i}a Civil, Universidad Michoacana de San Nicol\'as de Hidalgo. Edificio C, Ciudad Universitaria. Francisco J. M\'ujica s/n. Col. Fel\'icitas del R\'io. 58030, Morelia, Michoac\'an, M\'exico.\\
$^2$ Facultad de Ingenier\'ia, Universidad Aut\'onoma de Quer\'etaro, Cerro de las Campanas s/n, Colonia Las Campanas, Centro Universitario, 76010, Quer\'etaro, Quer\'etaro, M\'exico. \\
$^3$Instituto de F\'isica y Matem\'aticas, Universidad Michoacana de San Nicol\'as de
Hidalgo, Edificio C-3, Ciudad Universitaria.  Francisco J. M\'ujica s/n. Col. Fel\'icitas del R\'io.  58040 Morelia, Michoac\'an, M\'exico\\
$^4$Instituto de F\'isica, Pontificia Universidad Cat\'olica de Chile, Casilla 306, Santiago 22, Chile.
}
\ead{yconcha@umich.mx, adolfo.huet@uaq.mx, raya@ifm.umich.mx, devalenz@uc.cl}
\vspace{10pt}

\begin{abstract}
We study uniaxially strained graphene  under the influence of non-uniform magnetic fields perpendicular to the material sample with a coordinate independent strain tensor. For that purpose, we solve the Dirac equation with anisotropic Fermi velocity and  explore the conditions upon which such an equation possesses a supersymmetric structure in the quantum mechanical sense through examples. Working in a Laudau-like gauge, wave functions and energy eigenvalues are found analytically in terms of the magnetic field intensity, the anisotropy  scales and other relevant parameters that shape the magnetic field profiles.
\end{abstract}

%

\submitto{\JPCM}
%
\maketitle

\section{Introduction}

Modern material science, particularly connected with bidimensional materials (see, for instance~\cite{topins} for a recent review), has been boosted since the first isolation of graphene membranes~\cite{novo1}. This iconic material consist of a one-atom thick membrane of carbon atoms tightly packed in a honeycomb array, with outstanding properties for technological applications and fundamental physics development~\cite{rise}. Among other features, it is remarkable that the low-energy behavior of charge carriers in graphene is very much consistent with that of ultra-relativistic fermions inasmuch as the dispersion relation is linear. This implies that the equations of motion for these quasiparticles is a Dirac equation rather than the ordinary  Schr\"odinger equation with a typical parabolic dispersion relation. Theoretical studies of graphene date back to the seminal work of Wallace~\cite{Wallace} in which the pseudo-relativistic nature of charge carriers at low energy was pointed out. This is a first example of the nowadays abundant collection of Dirac-Weyl materials which open the possibility to connect high-energy physics dynamics in condensed matter systems  through the mathematical properties of the equations of motion. In particular, it is well known that the Dirac equation in some electromagnetic potentials 
can be factorized according to a supersymmetric structure in the quantum mechanical sense~\cite{susyqm1,susyqm2,susyqm3}.  This fact has been exploited to study  several properties of graphene~\cite{jorge,kuru,milpas,saul,kaushik,david1,david2} and other Dirac-like materials~\cite{kuru2,kuru3,yes} connected with the influence of external magnetic fields, as well as charge impurities which induce the atomic collapse effect when a supercritical regime is reached at~\cite{david}.
Interestingly, considering  a non-uniform magnetic field perpendicularly aligned to a membrane of graphene, in Ref.~\cite{kuru} the supersymmetry of the Dirac equation was exploited to study the magnetic states of charge carriers in this material for several profiles of the magnetic field which still permit an analytical solution of the said equation.

Because of the outstanding mechanical properties of monolayer graphene, namely its stiffness and strength~\cite{rise}, there has been an increasing interest to exploit strain to control other physical properties of the membranes. Straintronics~\cite{straintronics} has emerged as the study of mechanical deformations of graphene membranes to modify its properties (see Ref.~\cite{review} for a recent review). Response of graphene to tensile (positive) and compresive (negative) strain has been experimentally explored~\cite{tencomp1}. From the theoretical point of view, these mechanical deformation are encoded in a strain tensor which induces  a tensor character to the Fermi velocity in the material and consequently, the dispersion relation is modified from the ideal case and the low-energy regime corrects the equations of motion still in a tractable form~\cite{review}. A widely implemented idea to incorporate the influence of strain is through a pseudo-vector potential whose components are related to those of the position dependent strain tensor. This point of view preserves the Dirac nature of the equations of motion and explains in a natural manner the pseudo-Landau level observations in strained graphene, as predicted earlier on theoretical grounds~\cite{pLL}. The case of spatially uniform strain deserves special attention, since it is the limiting case of any general deformation. We consider this situation in the present article assuming a streching of the membrane. Moreover, the case is solvable and has a straightforward understanding by representing the lattice corrections in terms of a strained reciprocal space, which leads to an anisotropic Fermi velocity that does not produce any pseudo-magnetic field whatsoever~\cite{maurice}. Breaking of the isotropy of Fermi velocity is also observed in uniaxial strain~\cite{straintronics}, along with a displacement of the high symmetry Dirac points of the first Brillouin zone of the honeycomb array followed by a small tilting of the Dirac cones. All these observations have found a natural explanation in terms of an anisotropic variant of the Dirac equation with position dependent Fermi velocity~\cite{yonatan}. The resulting structure of the Dirac equation still allows for an analytical solution including the influence of an  external  uniform magnetic field~\cite{yonatan}, where uniaxial strain is seen to induce an contraction of the Landau levels spectra if the strain is either along the Zig-Zag or Arm-Chair edges of the graphene membrane.  
One of the goals of this article is to generalize the findings of~\cite{yonatan}, when uniform uniaxial strain is considered, under the influence of all magnetic field profiles considered in~\cite{kuru}. We explore the conditions upon which the supersymmetric structure of the anisotropic Dirac equation is preserved. We present the framework we choose to work in Sec.~\ref{sec:framework}. We work in detail the examples of a constant magnetic field, a trigonometric singular well, an exponentially decaying magnetic field, an hyperbolic singular magnetic field, a singular example and a hyperbolic well (or barrier) in Sec.~\ref{sec:examples}. Those examples are of theoretical interest in the sense that they lead to equations of motions that are analytically solvable. We finally conclude in Sec.~\ref{sec:conclu}.

\section{Framework}\label{sec:framework}
\subsection{General considerations}

Supersymmetry, or SUSY,  was introduced as a mathematical framework to unify the fundamental interactions in the celebrated standard model of particle physics. In this framework, a SUSY transformation relates fermions to bosons and viceversa~\cite{witten}. However, this symmetry is not observed. Therefore, it must have been broken at higher energies than the observed in particle accelerators so far. 
In order to study SUSY breaking in a simple settings, supersymmetric quantum mechanics (SUSY-QM) was introduced~\cite{susyqm1,susyqm2,susyqm3}. From an algebraic point of view, a quantum mechanical system described by a Hamiltonian ${\cal H}$ and $N$  operators $Q_i$, is said to be supersymmetric when the relation
\begin{equation}
    \{Q_i,Q_j^\dagger\}=Q_iQ_j^\dagger+Q_j^\dagger Q_i\equiv {\cal H}\delta_{ij}, \qquad i,j=1,\ldots N, \label{susyalgebra}
\end{equation}
holds. These operators are the so-called {\em supercharges} and the relation~(\ref{susyalgebra}) is a Lie superalgebra.

SUSY-QM has also a Hamiltonian realization. To present the general ideas of this framework, let us consider two Hamiltonians $H_1$ and $H_2$, each satisfying  a Schr\"odinger-like equation 
\begin{eqnarray}
\left[-\frac{d^2}{dx^2}+V_1(x) \right]\psi_1^n(x)\ \equiv \ H_1\psi_1^n&=&\varepsilon_{1}^n \psi_1^n(x)\;,\nonumber\\
\left[-\frac{d^2}{dx^2}+V_2(x)\right]\psi_2^n(x)\ \equiv \ H_2\psi_2^n&=&\varepsilon_{2} \psi_2^n(x)\;,\label{finalsusy}
\end{eqnarray}
where each of the potentials is  expressed in terms of a so-called superpotential $W(x)$ in the form
\begin{equation}
 V_1(x)=W^2(x)+W'(x)\;,\quad V_2(x)=W^2(x)-W'(x)\;,
\end{equation}
respectively. The potentials $V_1(x)$ and $V_2(x)$ are referred to as {\em supersymetric partner potentials}. Defining $ L^\pm=\mp \frac{d}{dx}+W(x)$,
one straightforwardly finds that these Hamiltonians can be factorized as $ H_1=L^-L^+$ and $ H_2=L^+L^-$, respectively. From this factorization, important properties follow.
To begin with, the eigenstates and eigenenergies of these two hamiltonians are related. Asumming the ground state energy of $H_1$ to vanish, $\varepsilon_1^0=0$, it is straightforward to verify that~\cite{susyqm1,susyqm2,susyqm3}
\begin{eqnarray}
\varepsilon_2^{n}&=&\varepsilon_1^{n+1},\nonumber\\
\psi_2^n&=& (\varepsilon_1^{n+1})^{-\frac{1}{2}} L^- \psi_1^{n+1},\nonumber\\
\psi_1^{n+1}&=& (\varepsilon_2^{n})^{-\frac{1}{2}} L^+ \psi_2^{n}.
\end{eqnarray}
Furthermore, by considering the matrix hamiltonian
\begin{equation}
    {\cal H}=\left(\begin{array}{cc} H_1 & 0 \\ 0 & H_2 \end{array} \right),
\end{equation}
and introducing the {\em supercharges}
\begin{equation}
    Q=\left(\begin{array}{cc} 0 & 0 \\ L^- & 0 \end{array} \right),\qquad 
    Q^\dagger=\left(\begin{array}{cc} 0 & L^+ \\ 0 & 0 \end{array} \right),
\end{equation}
it can be readily verified that these supercharges are conserved, namely, 
\begin{equation}
    [{\cal H},Q]=[{\cal H},Q^\dagger]=0,
\end{equation}
and that the following relations hold
\begin{equation}
    {\cal H}=\{Q,Q^\dagger\},\qquad \{Q,Q\}=\{Q^\dagger,Q^\dagger\}=0.
\end{equation}
So, the Hamiltonians $H_1$ and $H_2$ are said to be a supersymmetric quantum mechanical system. 
Within this framework, we consider the dynamics of graphene in a uniform magnetic field below.

\subsection{SUSY-QM and pristine graphene}
In this paper we are concerned with solving the 2-dimensional Dirac equation in the presence of a static magnetic field which is oriented perpendicular to the plane. It is well known that, even for an arbitrary spatial dependence of the magnetic field, this system can be factorized in the spirit of SUSY-QM in a pair  of equations corresponding to each component of the Dirac bi-spinor where each component would be under the influence of a superpartner potential. This means that the factorization of the system is the same regardless of the specifics of the magnetic field as long as it is static and perpendicular to the plane. This fact was exploited in a vast number of Refs.~\cite{kuru,david1,david2,kuru2,kuru3}. Particularly in Ref.~\cite{kuru}, the SUSY-QM scenario was exploited  to explore the spectra of graphene quasiparticles moving under the influence of several static, perpendicularly aligned magnetic fields with different spatial profiles. They commence with the stationary Dirac equation
\begin{equation}
    v_F \left( 
    \begin{array}{cc} 0 &  \Pi_x-i\Pi_y\\ \Pi_x+i\Pi_y & 0 \end{array}
    \right)\left(\begin{array}{c}\psi_1(x,y) \\ \psi_2(x,y)\end{array}\right) = E\left(\begin{array}{c}\psi_1(x,y) \\ \psi_2(x,y)\end{array}\right) \;,\label{Dirac}
\end{equation}
where $\vec\Pi=\vec p+e\vec A$, $\vec A$ represents the vector potential that describes the external magnetic field and $e$ the electric charge of these quasiparticles.
The resulting coupled system of equations for $\psi_1(x,y)$ and $\psi_2(x,y)$ can be decoupled in a standard manner, being equivalent to the uncoupled system of equations
\begin{eqnarray}
\Bigg(\Pi_x ^2-i[\Pi_x,\Pi_y]+\Pi_y^2 \Bigg)\psi_1   &=& \frac{E^2}{v_F^2}\psi_1\;, \nonumber\\
\Bigg(\Pi_x ^2+i[\Pi_x,\Pi_y]+ \Pi_y^2 \Bigg)\psi_2  &=& \frac{E^2}{v_F^2}\psi_2\;.
\label{decoupledini}
\end{eqnarray}
Working in a Landau-like gauge, we take
\begin{equation}
     A_x=0,\quad A_y=A_y(x),\quad [\Pi_x,\Pi_y]=-ie\hbar \frac{dA_y(x)}{dx}\equiv -ie\hbar B(x)\;.\label{scaling}
\end{equation}
Thus, defining 
\begin{equation}
     \varepsilon=\left(\frac{E}{\hbar v_F}\right)^2\;,\quad \psi_j(x,y)=e^{iky}\psi_j(x)\;, \quad j=1, 2,
\end{equation}
the system of equations in~(\ref{decoupledini}) can be written in the form
\begin{eqnarray}
\left[-\frac{d^2}{dx^2}+V_1(x) \right]\psi_1(x)\ \equiv \ H_1\psi_1&=&\varepsilon_{1} \psi_1(x)\;,\nonumber\\
\left[-\frac{d^2}{dx^2}+V_2(x)\right]\psi_2(x)\ \equiv \ H_2\psi_2&=&\varepsilon_{2} \psi_2(x)\;,\label{finalideal}
\end{eqnarray}
where
\begin{equation}
 V_{j}(x)=\left(k+\frac{eA_y(x)}{\hbar} \right)^2 + (-1)^{j-1} \frac{ eB(x)}{ \hbar}, \quad j=1, 2.
 \label{vunstrained}
\end{equation}
This happens to be equivalent to the system (\ref{finalsusy}) with
$W(x)=k+\frac{eA_y(x)}{\hbar}$.
 Our goal in this article is to verify the extent up to which the above structure is modified by assuming that the graphene membrane is subject to homogeneous uniaxial strain.

 \subsection{SUSY-QM and strained graphene}
 Graphene is an extraordinary material in many aspects. From its mechanical properties, it is remarkable that a membrane can support deformations of up to 20\% without breaking, modifying strongly the electronic properties of the material. Thus, the field of straintronics has grown a lot of interest in recent years~\cite{review}. Strain in graphene is introduced through the strain tensor $\epsilon$ (see, for instance, Ref.~\cite{straintronics}). For uniform deformations, the components of $\epsilon$ and coordinate independent, and thus its effect in the band structure of the material can be incorporated at the level of the tight-binding description~\cite{maurice}. For example, for a strain of $\epsilon_0$ percent along the Zig-Zag edge, we consider $\epsilon_{xx}=0.01\epsilon_0$, $\epsilon_{yy}=-\nu \epsilon_{xx}$ and $\epsilon_{xy}=0$, and we consider the Poisson ratio $\nu=0.1$, consistent with the rather low values of this quantity.  The displacement of carbon atoms due to strain affects also the hopping parameters and displace the position of the Dirac points with respect to the pristine sample and correspondingly, the dispersion relation is corrected in such a manner that in the low energy regime, the Fermi velocity acquires a tensor, anisotropic nature such that the isoenergetic contours are no longer circles around the Dirac points, but ellipses with an eccentricity dictated by the ratio of elongation along the Zig-Zag and Armchair edges of the honeycomb array induced by the strain.  
%
%
 Under these considerations,  
the corresponding stationary Dirac equation reads
\begin{equation}
    v_F \left( 
    \begin{array}{cc} 0 & a \Pi_x-ib\Pi_y\\ a\Pi_x+ib\Pi_y & 0 \end{array}
    \right)\left(\begin{array}{c}\psi_1(x,y) \\ \psi_2(x,y)\end{array}\right) = E\left(\begin{array}{c}\psi_1(x,y) \\ \psi_2(x,y)\end{array}\right) \;,
    \label{hamstrainin}
\end{equation}
where the quantities $\Pi_{x,y}$ and $A_{x,y}$ are the same as in the non-strained, regular case, but $a$ and $b$ accounting for the deformation along the edges of graphene. 
These functions are merely constants accounting for small deformations of graphene membranes, and correspond  to a limiting case of more general uniaxial strains that can be considered. From Eq.~(\ref{hamstrainin}), the resulting coupled system of equations, through a standard procedure, can be expressed as
\begin{eqnarray}
\Bigg(a^2 \Pi_x ^2-iab[\Pi_x,\Pi_y]+b^2 \Pi_y^2 \Bigg)\psi_1 &=& \frac{E^2}{v_F^2}\psi_1 \nonumber\\
\Bigg(a^2 \Pi_x ^2+iab[\Pi_x,\Pi_y]+b^2 \Pi_y^2 \Bigg)\psi_2 &=& \frac{E^2}{v_F^2}\psi_2\;.
\label{decoupledstrain}
\end{eqnarray}
Thus, by adopting the Landau-like gauge to express the external magnetic field and defining
\begin{equation}
      \psi_j(x,y)=e^{iky/\zeta}\psi_j(x)\;, \quad\zeta=\frac{b}{a},\quad j=1, 2,
\end{equation}
the system of equations~(\ref{decoupledstrain}) is equivalent to
\begin{eqnarray}
\left[-\frac{d^2}{dx^2}+V_1^{\zeta}(x) \right]\psi_1(x)&=&\varepsilon_{a,1} \psi_1(x)\;,\nonumber\\
\left[-\frac{d^2}{dx^2}+V_2^{\zeta}(x)\right]\psi_2(x)&=&\varepsilon_{a,2} \psi_2(x)\;.\label{final}
\end{eqnarray}
where 
\begin{equation}
 V_{j}^{\zeta}(x)=\zeta^2\left(k+\frac{eA_y(x)}{\hbar} \right)^2 + (-1)^{j-1} \zeta\frac{ eB(x)}{ \hbar}, \quad \varepsilon_{a,j}=\frac{\varepsilon_j}{a^2}\;,\quad j=1, 2.\label{vstrain}
\end{equation}
 The case $\zeta=1$ corresponds to the ideal case, whereas $\zeta<1$ dictates that strain is applied along the Zig-Zag edge, and $\zeta>1$ along the Armchair edge.
One can observe that, just as in the ideal case, the system of Eqs.~(\ref{final}) has been decoupled into a supersymmetric system. We further note that the effect of strain in the potential is  solely parametrized by $\zeta$. Furthermore, upon considering the profile of the external magnetic field to be given in terms of the vector potential $A_y(x)$ as
\begin{equation}
A_y(x)=\frac{B_0}{\alpha}F(\alpha x)\;,\qquad
    \vec{B}(x)=\left(0,0,B_0 \frac{d}{dx}F(\alpha x)\right),
\end{equation}
with $B_0$ constant and $F$ some smooth function, the system of Eqs.~(\ref{final}) can be cast in the explicit form
\begin{eqnarray}
    \left[-\frac{d^2}{dx^2}-\frac{d}{dx}\left( k+\frac{\beta}{\alpha}F(\alpha x)\right) + \left( k+\frac{\beta}{\alpha}F(\alpha s)\right)^2 \right]\psi_1(x)&=&\varepsilon_{a,1}\psi_1(x)\;,\nonumber\\
    \left[-\frac{d^2}{dx^2}+\frac{d}{dx}\left( k+\frac{\beta}{\alpha}F(\alpha x)\right) + \left( k+\frac{\beta}{\alpha}F(\alpha s)\right)^2 \right]\psi_2(x)&=&\varepsilon_{a,2}\psi_2(x)\;,
\end{eqnarray}
with $\beta=\zeta e B_0/\hbar$,
which is identical to Eqs.~(\ref{finalideal}) and~(\ref{vunstrained}),  provided we identify
\begin{equation}
    B_0\to \zeta B_0,\qquad k\to k/\zeta ,\qquad \varepsilon_{j}\to\varepsilon_{a,j}. \label{replacements}
\end{equation}
Below we exploit this identification to explore the role of strain in the energy eigenvalues corresponding to several examples worked in literature~\cite{kuru}.


\section{The role of strain through examples}\label{sec:examples}

In this section we revisit the examples considered in Ref.~\cite{kuru} of the solution of the 2-dimensional Dirac equation under several static but non-uniform magnetic fields, with the added ingredient of (tensile) strain. As we have seen above, this can be accomplished in a direct manner from the ideal, unstrained case through the replacements~(\ref{replacements}).  We particularly focus on the impact of the parameters $\zeta$ (or independently on $a$, and $b$) on the energy eigenvalues for a given magnetic field profile.

\subsection{The constant magnetic field}

\begin{figure}
\begin{center}
\includegraphics[width=0.5\textwidth]{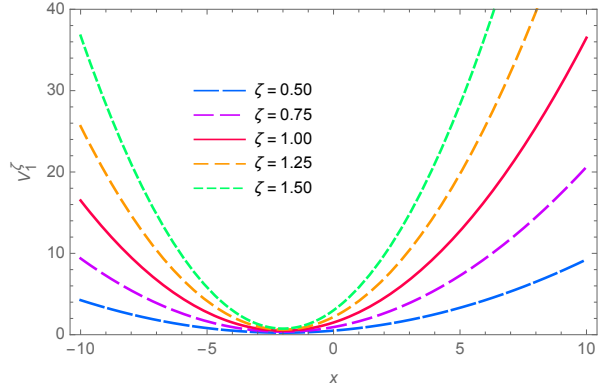}
\caption{Variation of $V_1(x)$ for the constant magnetic field with respect to $\zeta$. We have set $D=0.5$ and $k=1$.}
\label{V1cmf}
\end{center}
\end{figure}

We first consider the case of a uniform magnetic field, expressed through
\begin{equation}
A_y=B_0 x,\qquad \vec{B}=\left(0,0,B_0 \right),
\end{equation}
which is known to give rise to the Landau levels~\cite{yonatan}. The corresponding super partner potentials, according to Eq.~(\ref{vstrain}), are
\begin{equation}
 V_{j}^{\zeta}(x)=\zeta^2\left(k+Dx \right)^2 + (-1)^{j-1} \zeta D, 
 \quad j=1, 2,
\end{equation}
with $D= eB_0/\hbar$.
It corresponds to a displaced harmonic oscillator potential and is shown in Fig.~\ref{V1cmf} for different values of the strain parameter $\zeta$. 
From these potentials it follows that the spectrum is
\begin{equation}
\varepsilon_{2,a}^0=0\,, \qquad  \varepsilon_{2,a}^n=\varepsilon_{1,a}^{n-1}= \zeta (2D) n \;,\quad  n=1,2,3\ldots 
\end{equation}
or
\begin{equation}
E_{2,a}^0=0\,, \qquad  E_{2,a}^n=E_{1,a}^{n-1}= \hbar v_F \sqrt{a b (2D) n} \;,\quad  n=1,2,3\ldots
\end{equation}
As in the non-strained case~\cite{yonatan}, the energy levels are directly dependent on $D$ but do not depend on $k$. In this case, increasing $a$ or $b$  increase the energy levels.


\subsection{Hyperbolic well or barrier}

\begin{figure}
\begin{center}
\includegraphics[width=0.5\textwidth]{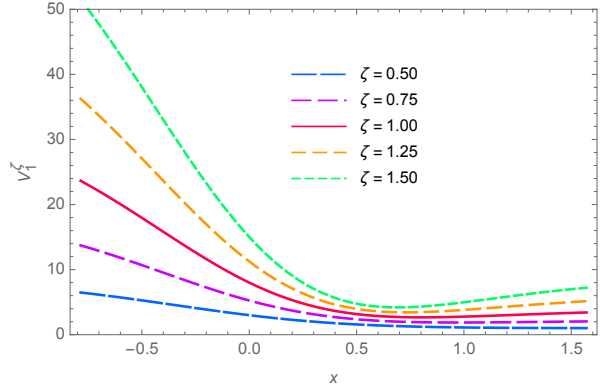}
\caption{Variation of $V_1(x)$ for the hyperbolic well with respect to $\zeta$. We have set $D=4$ and $k=-2$.}
\label{Vhwell}
\end{center}
\end{figure}

The magnetic field for this case is specified through 
\begin{equation}
A_{y}(x)=-\frac{B_{0}}{\alpha}\tanh \alpha x \;, \qquad \vec{B}(x)=\left(0,0,\frac{B_0}{\cosh^2 \alpha x}\right)\,.
\end{equation}
To appreciate better the effect of strain,  we set $\alpha=1$.
Then, the superpartner potentials ($V_1$ is shown in Fig.~\ref{Vhwell}) are now
\begin{equation}
    V_{j}^{\zeta}(x) = \zeta^2 (k^2+D^2) +2k\zeta^2 D \tanh(x) -\zeta D(\zeta D + (-1)^{j}) \mbox{sech}^2(x)\;,
\end{equation}
For this type of magnetic field, Eqs.~(\ref{final}) assume the form 
\begin{equation}
\left[-\frac{d^2}{d x^2} + \zeta^2 (k^2+D^2) +2k\zeta^2 D \tanh(x) -\zeta D(\zeta D + (-1)^{j}) \mbox{sech}^2(x) - \varepsilon_a\right]\psi_{j}(x)=0\;,
\end{equation}
with
 \begin{equation}
 A\equiv \zeta D,\quad B\equiv \zeta^2 k D, \quad \varepsilon_{ab}\equiv \varepsilon_a-\zeta^2( k^2 + D^2)\;,
  \end{equation}
obtaining the system
\begin{equation}
\left[\frac{d^2}{d x^2} - 2B \tanh(x) +A(A + (-1)^{j}) \mbox{sech}^2(x) - \varepsilon_{ab}\right]\psi_{j}(x)=0\;.
\end{equation}  
Writing the above expression  in a Sturm-Liouville form,  we readily obtain the  spectrum
\begin{equation}
    \varepsilon_{1,ab}^n = -(A-n)^2 - \frac{B^2}{(A-n)^2}\;,\quad
    \varepsilon_{2,ab}^{n}=\varepsilon_{1,ab}^{n-1} 
    \;,\quad  n=1,2,3\ldots 
\end{equation}
Then,
\begin{equation}
    \varepsilon_{1,a}^n = \zeta^2(k^2 + D^2) -(\zeta D-n)^2 - \frac{\zeta^4 k^2 D^2}{(\zeta D-n)^2}\;,\quad
    \varepsilon_{2,a}^{n}=\varepsilon_{1,a}^{n-1} 
    \;,\quad  n=1,2,3\ldots 
\end{equation}
Taking $u=\tanh(x)$, $s_1=\zeta D -1$, $s_2=s_1 + 1$, $a_1=\frac{\zeta^2 kD}{\zeta D - n+1}$ and $a_2=\frac{\zeta^2 k D}{\zeta D - n}$, the corresponding eigenfunctions are
\begin{equation}
\psi_{j}^n(u(x))=(1-u)^{(s_j - n + a_j)/2}(1 + u)^{(s_j - n - a_j)/2}P_n^{(s_j-n+a_j,s_j-n-a_j)}(u)\;, \quad j=1,2
\end{equation}
where $P_{n}^{(\alpha,\beta)}(u)$ are the Jacobi polynomials beging $\alpha, \beta >1$.
Finally, the energy eigenvalues are 
\begin{eqnarray}
E_{1}^n&=&\hbar v_F\sqrt{b^2(k^2 + D^2) -(bD - a n)^2 -\frac{b^4 k^2 D^2}{(bD - an)^2}}\;,
\\
E_{2}^n&=&\hbar v_F\sqrt{b^2(k^2 + D^2) -(bD - a(n -1))^2 -\frac{b^4 k^2 D^2}{(bD - a(n -1))^2}}\;,
\end{eqnarray}
for integer $n$ and starting from $n=1$ which implies that $E_{2}^0=0$

We note that, for a fixed $k$,  the ratio of $D$ and $a$ determines the number of discrete levels, meaning that this number is strain dependent.
If we fix these  parameters, the number of levels is increased or decreased by $k$ but its  role  can be either diminished or enhanced by the strain parameter $b$.

\subsection{The trigonometric singular well}
\begin{figure}
\begin{center}
\includegraphics[width=0.5\textwidth]{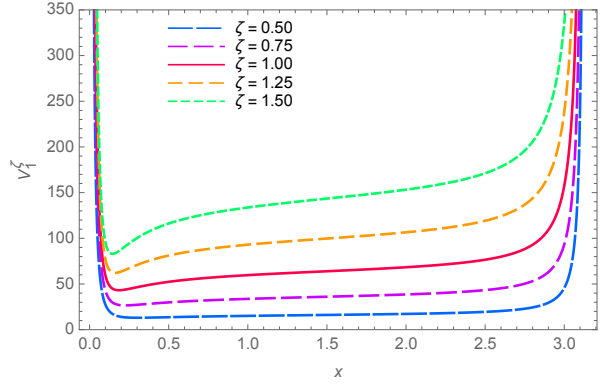}
\caption{Variation of $V_1(x)$ for the trigonometric singular well with respect to $\zeta$. We have set $D=0.5$ and $k=8$.}
\label{V1VSz}
\end{center}
\end{figure}

We analyze the magnetic field  obtained from the potential 
\begin{equation}
    A_y(x)=-\frac{B_0}{\alpha}{\rm cot}(\alpha x)\;, 
    \qquad \vec{B}(x)=\left(0, 0, \frac{B_0}{{\rm cot}^2(\alpha x)} \right)\,,
\end{equation}
under uniform strain parametrized by $\zeta$.
We again set $\alpha=1$. This leads to the potentials
\begin{equation}
    V_{j}^{\zeta}(x) = \zeta^2 (k^2-D^2)
    - \zeta^2 2kD {\rm cot}(x) 
    + \zeta D\left( \zeta D + (-1)^{j-1} \right){\rm csc}^2(x)\;,
\end{equation}
where $D=\frac{e B_0}{\hbar }$.
These potentials are infinite wells that restrict the problem to the interval
$0 < x < \pi$.
The effect of the strain parameter $\zeta$ on the potential is shown in Fig.~\ref{V1VSz}.
For this type of magnetic field, Eqs.~(\ref{final}) assume the form 
\begin{equation}
\left[-\frac{d^2}{dx^2}+\zeta^2 (k^2-D^2)
    - \zeta^2 2kD {\rm cot}(x) 
    + \zeta D\left( \zeta D \pm 1 \right){\rm csc}^2(x) 
    -\varepsilon_a \right]\psi_{1,2}(x) = 0 \;.
\end{equation}
 Introducing the notation
\begin{equation}
 A\equiv \zeta D,\quad B\equiv \zeta^2 k D, \quad \varepsilon_{ab}\equiv \varepsilon_a+ \zeta^2 (D^2-k^2)
\end{equation}
we obtain the system 
\begin{equation}
 \left[\frac{d^2}{dx^2}
    +  2B {\rm cot}(x) 
    - A\left( A \pm 1 \right){\rm csc}^2(x) 
    -\varepsilon_{ab} \right]\psi_{1,2}(x) = 0,   
\end{equation}
which can be cast in a Sturm-Liouville form and solved following the method used in~\cite{Kirchbach}. We obtain the spectrum
\begin{equation}
    \varepsilon_{1,ab}^n = (A+n)^2 - \frac{B^2}{(A+n)^2}\;,\quad
    \varepsilon_{2,ab}^{n}=\varepsilon_{1,ab}^{n-1} 
    \;,\quad  n=1,2,3\ldots \\
\end{equation}
which finally results in
\begin{eqnarray}
    E_{1,n} &=& \hbar v_F  \sqrt{b^2(k^2-D^2) + (bD + an)^2
    -\frac{b^4 k^2 D^2}{(bD + an)^2} }\;, \\
    E_{2,n} &=& \hbar v_F  \sqrt{b^2(k^2-D^2) + (bD + a(n-1))^2
    -\frac{b^4 k^2 D^2}{(bD + a(n-1))^2} }\;,
  \end{eqnarray}
for integer $n$ and starting from $n=1$, which implies that $E_{2,0}=0$.

We observe that, regardless of strain, this potential produces an infinite number of bound states for any real value of $k$.
The effect of strain is that the energy eigenvalues (which are positive) get increased as $a$ or $b$ become larger than one. Conversely, the energies are lowered when any of these parameters is smaller than one.



\subsection{Exponentially decaying magnetic field}
To define the exponentially decaying magnetic field, let us take 
\begin{equation}
    A_y(x)=-\frac{B_0}{\alpha}(e^{-\alpha x}-1)\;, \qquad B(x)=B_0e^{-\alpha x}\,.\label{expo}
\end{equation}
This choice of $A_y$ ensures that for $\alpha=0$ we recover the constant field results, as opposed to the gauge choice of Ref.~\cite{kuru}. Defining $D=eB_0/(\alpha\hbar)$, the potentials have the form
\begin{equation}
    V_j^\zeta(x)= \zeta^2 (D+k)^2 + \zeta^2 D^2 e^{-2\alpha x} +\zeta \left( (-1)^{j-1}\alpha - 2\zeta(k+D)\right)De^{-\alpha x}.
\end{equation}
For this field configuration, Eqs.~(\ref{final}) are equivalent to
\begin{equation}
\left[-\frac{d^2}{dx^2}+\zeta^2(k+D)^2-\varepsilon_a+\zeta \left((-1)^{j-1} \alpha-2\zeta (k+D)\right)De^{-\alpha x}+\zeta^2 D^2e^{-2\alpha x} \right]\psi_j(x)=0.
\end{equation}
Let $u=(2D\zeta/\alpha)e^{-\alpha x}$. Then, the above equation can be cast in the form
\begin{equation}
    \left[ \frac{d^2}{du^2}+\frac{1}{u}\frac{d}{du}-\frac{s^2}{u^2}+\frac{\nu_j}{2u}-\frac{1}{4}\right]\psi_j=0\;,\label{psib}
\end{equation}
where
\begin{equation}
    s^2=\frac{1}{\alpha^2}\left(\zeta^2(k+D)^2-\varepsilon_a \right)\;, \qquad \nu_j= \frac{2\zeta}{\alpha }(k+D)+(-1)^{j-1}.
\end{equation}
With the ansatz
\begin{equation}
    \psi_B=e^{-u/2}u^s F_j(u)\;,
\end{equation}
either of Eqs. in~(\ref{psib}) is equivalent to
\begin{equation}
    u \frac{d^2F_j}{du^2}+(2s+1-u)\frac{dF_j}{du}-\left( s+\frac{1-\nu_j}{2}\right) F_j=0,
\end{equation}
which has as solution the confluent hypergeometric function
\begin{equation}
    F_j(u)=_1F_1\left( s+\frac{1-\nu_j}{2}; 2s+1; u\right)\;,
\end{equation}
which in order to be normalizable demands that
\begin{equation}
     s+\frac{1-\nu_j}{2}=-n,\qquad n=0, 1, 2, \ldots
\end{equation}
Thus, the eigenvalues in this case are
\begin{equation}
    \varepsilon_{1,a}^n=\zeta^2(k+D)^2-(-\alpha(n+1)+\zeta(k+D))^2=\varepsilon_{2,a}^{n-1},
\end{equation}
 and correspondingly,
\begin{eqnarray}
        E_1^n&=&\hbar v_F \sqrt{b^2(k+D)^2-\left(-\alpha a n + b (k+D) \right)^2}\;,\nonumber\\
        E_2^n&=&\hbar v_F \sqrt{b^2(k+D)^2-\left(-\alpha a (n-1) + b (k+D) \right)^2}\;.
\end{eqnarray}
In the limit $\alpha=0$,
\begin{equation}
    E_1^{n-1} \Bigg|_{\alpha=0}=\hbar v_F \sqrt{2 a b B_0 n)}=E_2^{n}\Bigg|_{\alpha=0},
\end{equation}
in agreement with the eigenvalues for the uniform magnetic field case.

\begin{figure}
\begin{center}
\includegraphics[width=0.5\textwidth]{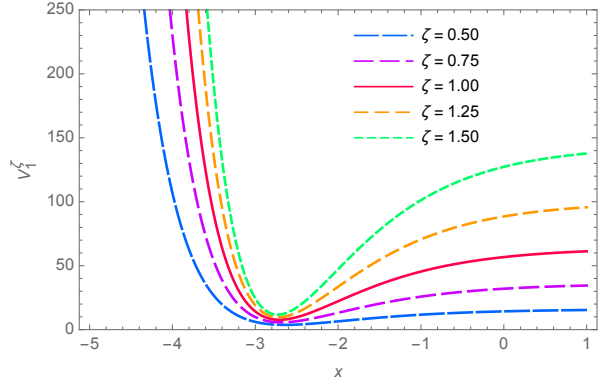}
\caption{Variation of $V_1(x)$ for the exponential decaying magnetic field with respect to $\zeta$. We have set $D=0.5$ and $k=8$.}
\label{V1edmf}
\end{center}
\end{figure}


We observe that, as in the non-strained case, the allowed number of energies in the discrete spectrum depends $k$, but now it is also controlled by the $\zeta$ parameter. Moreover, the choice of gauge has made the energy eigenvalues explicitly dependent on the intensity of the magnetic field given through $D$. This is in contrast with the gauge choice of Ref.~\cite{kuru}


\subsection{Hyperbolic singular field}

\begin{figure}
\begin{center}
\includegraphics[width=0.5\textwidth]{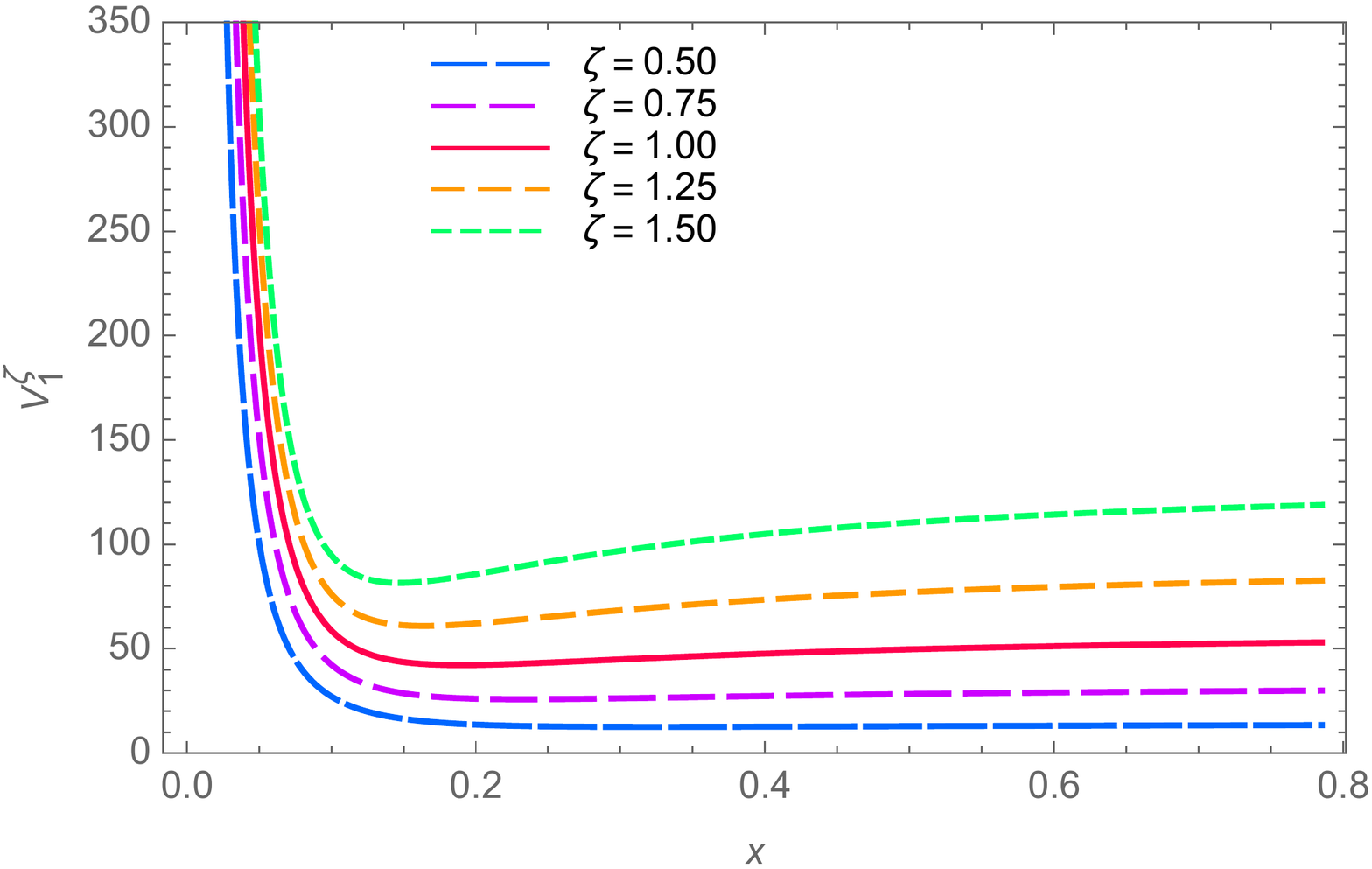}
\caption{Variation of $V_1(x)$ for the hyperbolic singular field with respect to $\zeta$. We have set $D=0.5$ and $k=8$.}
\label{V1hsf}
\end{center}
\end{figure}

The magnetic field for this case is described though
\begin{equation}
A_{y}(x)=-\frac{\hbar D}{e}\coth \alpha x \;, \qquad \vec{B}(x)=\left(0,0,\frac{B_0}{\sinh^2 \alpha x}\right)\,.
\end{equation}
To appreciate better the effect of strain, again we set $\alpha=1$.
Then, the superpartner potentials are
\begin{equation}
    V_{j}^{\zeta}(x) = \zeta^2 (k^2+D^2) -2k\zeta^2 D \coth(x) +\zeta D(\zeta D +  (-1)^{j-1}) \mbox{cosech}^2(x)\;,
\end{equation}
For this type of magnetic field, eqs.~(\ref{final}) assume the form 
\begin{equation}
\left[-\frac{d^2}{d x^2} + \zeta^2 (k^2+D^2) -2k\zeta^2 D \coth(x) +\zeta D(\zeta D +  (-1)^{j-1}) \mbox{cosech}^2(x) - \varepsilon_a\right]\psi_{j}(x)=0\;,\label{hsf}
\end{equation}
with
 \begin{equation}
 A\equiv \zeta D,\quad B\equiv \zeta^2 k D, \quad \varepsilon_{ab}\equiv \varepsilon_a-\zeta^2( k^2 + D^2)\;,
  \end{equation}
we re-write~(\ref{hsf}) as
\begin{equation}
\left[\frac{d^2}{d x^2} + 2B \coth(x) -A(A+ (-1)^{j-1}) \mbox{cosech}^2(x) - \varepsilon_{ab}\right]\psi_{j}(x)=0\;,
\end{equation}  
which already has the Sturm-Liouville form, from where we obtain the spectrum
\begin{equation}
    \varepsilon_{1,ab}^n = -(A+n)^2 - \frac{B^2}{(A+n)^2}\;,\quad
    \varepsilon_{2,ab}^{n}=\varepsilon_{1,ab}^{n-1} 
    \;,\quad  n=1,2,3\ldots ,
\end{equation}
or, in the original parameters
\begin{equation}
    \varepsilon_{1,a}^n = \zeta^2(k^2 + D^2) -(\zeta D+n)^2 - \frac{\zeta^4 k^2 D^2}{(\zeta D+n)^2}\;,\quad
    \varepsilon_{2,a}^{n}=\varepsilon_{1,a}^{n-1} 
    \;,\quad  n=1,2,3\ldots \\
\end{equation}
Their associated eigenfunctions take the form
\begin{equation}
\psi_{j}^n(u(x))=(u -1)^{-(s_j + n - a_j)/2}(u + 1)^{-(s_j + n + a_j)/2}P_n^{(-s_j-n+a_j,-s_j-n-a_j)}(u)\;, \quad j=1,2
\end{equation}
with $u=\coth(x)$, $s_2=\zeta D$, $a_2=\frac{\zeta^2 kD}{(\zeta D + n)}$, $s_1=s_2 + 1$ and $a_1=\frac{\zeta^2 k D}{(\zeta D + n + 1)}$, where $P_n^{(\alpha,\beta)}(u)$ are Jacobi polynomials with $\alpha, \beta>-1$. Finally, the energy eigenvalues are 
\begin{eqnarray}
E_{1}^n&=&\hbar v_F\sqrt{b^2(k^2 + D^2) -(bD + a n)^2 -\frac{b^4 k^2 D^2}{(bD + an)^2}}\;,\\
E_{2}^n&=&\hbar v_F\sqrt{b^2(k^2 + D^2) -(bD + a(n -1))^2 -\frac{b^4 k^2 D^2}{(bD + a(n -1))^2}}\;,
\end{eqnarray}
for integer $n$ and starting from $n=1$, which implies $E_{2}^0=0$

For this potential, we note that regardless of strain, the number of levels depends on $k$ but now it also depends on the strain parameter $b$.

\subsection{Singular magnetic field}

\begin{figure}
\begin{center}
\includegraphics[width=0.5\textwidth]{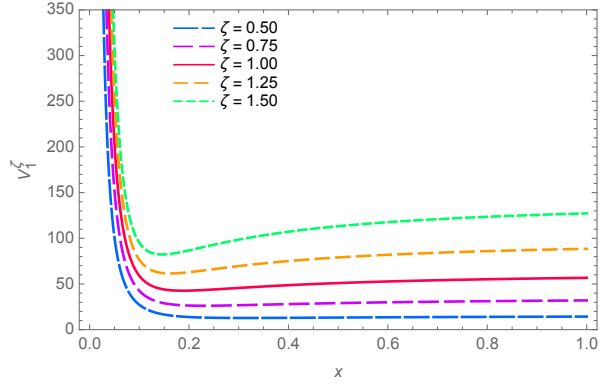}
\caption{Variation of $V_1(x)$ for the singular magnetic field with respect to $\zeta$. We have set $D=0.5$ and $k=8$.}
\label{V1smf}
\end{center}
\end{figure}

We describe magnetic field for this case as
\begin{equation}
A_{y}(x)=-\frac{B_{0}}{x}\;, \qquad \vec{B}(x)=\left(0,0,\frac{B_0}{x^2}\right)\,.
\end{equation}
Now, the superpartner potentials are
\begin{equation}
    V_{j}^{\zeta}(x) = \zeta^2 k^2 +\frac{\zeta D(\zeta D + (-1)^{j-1})}{x^2}
-\frac{2k\zeta^2 D}{x}\;,
\end{equation}
where $D=\frac{e B_0}{\hbar}$.
These potentials diverges in the origin but vanish as $x \to + \infty$. The effect of the strain parameter on these potentials is shown in  Fig.~\ref{V1smf}.  For this type of magnetic field, Eqs.~(\ref{final}) take the form
\begin{equation}
\left[-\frac{d^2}{dx^2}+\zeta^2 k^2
    - \frac{2\zeta^2 kD}{x}  
    + \frac{\zeta D\left( \zeta D + (-1)^{j-1} \right)}{x^2} 
    -\varepsilon_a \right]\psi_{j}(x) = 0 \;.\label{ec. singular magnetic}
\end{equation}
 Introducing  the notation
\begin{equation}
 A\equiv \zeta D,\quad B\equiv \zeta^2 k D, \quad \varepsilon_{ab}\equiv \varepsilon_a-\zeta^2 k^2
\end{equation}
we obtain
\begin{equation}
 \left[\frac{d^2}{dx^2}
    +  \frac{2B}{x} 
    - \frac{A\left( A + (-1)^{j-1} \right)}{x^2}
    -\varepsilon_{ab} \right]\psi_{j}(x) = 0   
\end{equation}
The eigenvalues of this problem are well known, 
\begin{equation}
 \varepsilon_{1,ab}^n =-\frac{B^2}{(n + A)^2}\; , \quad
     \varepsilon_{2,ab}^{n}=\varepsilon_{1,ab}^{n-1} 
     \;,\quad  n=1,2,3\ldots \\
\end{equation}
Then
\begin{equation}
\varepsilon_{1,a}^{n}=\zeta^4 k^2D^2\left( \frac{1}{\zeta^2D^2} - \frac{1}{(n + \zeta D)^2}\right)\;, \quad \varepsilon_{2,a}^{n}=\varepsilon_{1,a}^{n-1} 
    \;,\quad  n=1,2,3\ldots 
\end{equation}
The eigenfunctions can be expressed in terms of Laguerre polynomials $L^{\alpha}_{n}(x)$: 
\begin{equation}
\psi_{1}^{n}(u_{1}(x))=u_{1}^{\zeta D + 1}e^{-u_{1}/2}L_{n}^{2\zeta D + 1}(u_{1})\;,  \qquad  u_{1}=\frac{2\zeta^2 kD}{n + \zeta D + 1}x,
\end{equation}
and
\begin{equation}
\psi_{2}^{n}(u_2(x))=u_{2}^{\zeta D}e^{-u_{2}/2}L_{n}^{2\zeta D - 1}(u_2)\;, \qquad  u_{2}=\frac{2\zeta^2 kD}{n + \zeta D}x\;.
\end{equation}
The  condition $D>1$ guarantees the square-integrability of these eigenfuntions. Then, the energy eigenvalues of the equation (\ref{ec. singular magnetic}) are
\begin{eqnarray}
E_{1}^n&=&\hbar v_{F}b^2kD\sqrt{\frac{1}{(bD)^2} -\frac{1}{(an +bD)^2}}\;,\\
E_{2}^n&=&\hbar v_{F}b^2kD\sqrt{\frac{1}{(bD)^2} -\frac{1}{(a(n-1) +bD)^2}}\;,
\end{eqnarray}
for integer $n$ and starting from $n=1$ which implies that $E_{2}^0=0$.

\section{Conclusions}~\label{sec:conclu}
In this work, we have studied electronic states in  uniformly strained graphene under the influence of inhomogeneous magnetic fields pointing in the transverse direction to the plane of the membrane, with an explicit dependence on its coordinates. To this end, we have taken into account anisotropic corrections to the Fermi velocity which still render the 2-dimensional Dirac equation in a tractable form. The supersymmetric quantum mechanical structure of this equation allows to solve it in those cases of inhomogeneous magnetic fields available in literature~\cite{kuru}. The effect of strain at the level of the super-partner potentials is solely  encoded in the parameter $\zeta$ which measures the ratio of strain coefficients $b$ and $a$ along each spatial dimension, respectively.  With respect to the energy levels, our analysis shows that the effects of strain do depend on $a$ and $b$ separately in general and that each of the parameters may have a different effect depending on the type of potential. It is only in the constant magnetic field that the field intensity  gets normalized by $B_0\to \sqrt{ab}B_0$ in the Laudau-level spectra~\cite{yonatan}. Moreover, we have shown that for some potentials  that allow a finite number of energy levels in the discrete spectrum, the strain can play a role in determining the number of levels. All our analysis was done explicitly for the magnetic field cases discussed in Ref.~\cite{kuru}. For the exponentially decaying magnetic field case, we demonstrate that  a careful gauge choice is important to obtain the constant magnetic field limit when the damping factor $\alpha$ in Eq.~(\ref{expo}) vanishes. Our findings correspond to a limiting case of more general coordinate dependent uniaxial strain. Moreover, limitations of our approach can be explored in further detail can be explored in further detail following, for instance, Ref.~\cite{yonatan}. In particular, the evolution  of the geometrical parameters $a$ and $b$ (or $\zeta$) with varying the Poisson ratio and tensile strain must be explored in further detail, which is now under study along with its compresive counterpart and results shall be discussed elsewhere.

\ack
YC acknowledges support from CIC-UMSNH under grant 3289384. AR acknowledges support from Consejo Nacional de Ciencia y Tecnolog\'ia (M\'exico) under grant 256494.

\end{document}